\def\be{\begin{equation}}
\def\te{\end{equation}}
\def\ba{\begin{eqnarray}}

\def\ta{\end{eqnarray}}

\def\be{\begin{equation}}
\def\ee{\end{equation}}
\def\bea{\begin{eqnarray}}

\def\eea{\end{eqnarray}}

\newskip\humongous \humongous=0pt plus 1000pt minus 1000pt

\newif\ifdtup

\def\(#1){(\ref{#1})}

\documentstyle[prd,aps,eqsecnum]{revtex}

\begin{document}
\draft
\title{A Kinetic Theory Approach to Quantum Gravity}
\author{B. L. Hu}
\address{Department of Physics,
         University of Maryland,
         College Park, MD 20742, USA.
         hub@physics.umd.edu\\
-- Invited talk given at the 6th Peyresq Meeting, France, June,
2001. To appear in Int. J. Theor. Phys.}
\date{03/27/2002. Report umdpp \#02-039}
\maketitle
\begin{abstract}
We describe a kinetic theory approach to quantum gravity --  by
which we mean a theory of the microscopic structure of spacetime,
not a theory obtained by quantizing general relativity. A
figurative conception of this program is like building a ladder
with two knotted poles: quantum matter field on the right and
spacetime on the left. Each rung connecting the corresponding
knots represent a distinct level of structure. The lowest rung is
hydrodynamics and general relativity; the next rung is
semiclassical gravity, with the expectation value of quantum
fields acting as source in the semiclassical Einstein equation.
We recall how ideas from the statistical mechanics of interacting
quantum fields helped us identify the existence of noise in the
matter field and its effect on metric fluctuations, leading to
the establishment of the third rung: stochastic gravity,
described by the Einstein-Langevin equation. Our pathway from
stochastic to quantum gravity is via the correlation hierarchy of
noise and induced metric fluctuations. Three essential tasks
beckon: 1) Deduce the correlations of metric fluctuations from
correlation noise in the matter field; 2) Reconstituting quantum
coherence -- this is the reverse of decoherence -- from these
correlation functions 3) Use the Boltzmann-Langevin equations to
identify distinct collective variables depicting recognizable
metastable structures in the kinetic and hydrodynamic regimes of
quantum matter fields and how they demand of their corresponding
spacetime counterparts. This will give us a hierarchy of
generalized stochastic equations -- call them the
Boltzmann-Einstein hierarchy of quantum gravity -- for each level
of spacetime structure, from the macroscopic (general relativity)
through the mesoscopic (stochastic gravity) to the microscopic
(quantum gravity).
\end{abstract}
\baselineskip=15pt 

\newpage
\section{Introduction}
\label{sec:intro}


In the last decade  a statistical mechanics description of
particles, fields and spacetime based on the concept of quantum
open systems and the influence functional formalism has been
introduced. It reproduces in full the established theory of
quantum fields in curved spacetime
\cite{DeW75,BirDav,Fulling,Wald,Mostepanenko,MirVil} and contains
also a microscopic description of their stochastic properties,
such as noise, fluctuations, decoherence, and dissipation. This
new framework allows one to explore the quantum statistical
properties of spacetime beyond the semiclassical regime, as well
as important non-equilibrium processes in the early universe and
black holes, such as particle creation, entropy generation,
structure formation, Hawking radiation, horizon fluctuations,
backreaction and the black hole information `loss' issues. This
theory describing particles and fields is known as statistical,
stochastic or kinetic field theory, while that applied to
spacetime dynamics defines the stochastic gravity program.
\footnote{One can find a sample of original papers leading to the
establishment of this field in \cite{Physica,ELE,MV012,HP012} and
recent reviews in \cite{stogra,HVErice} (the first is mainly on
ideas, the second is technical). On statistical field theory, one
can find the material most relevant to our discussions here,
i.e., the correlation hierarchy, master effective action,
decoherence of correlation history and stochastic Boltzmann
equation, in \cite{CH88,dch,cddn,stobol}.}

The search on the gravity side proceeded from the classical level
described by Einstein's general relativity theory to the
semiclassical level described by the semiclassical Einstein
equation. (This is semiclassical gravity, or curved spacetime
quantum field theory with backreaction \cite{CH87,CV94}.)
Stochastic gravity is at the next higher level.  The progression
on the matter side is better known. The classical matter is
usually described by a hydrodynamic equation of state. At the
semiclassical level it is given by the expectation value of the
energy momentum tensor operator of matter fields with respect to
some quantum state. We have the microscopic theory of ordinary
matter -- QED and QCD -- so it is easy to deduce its meso and
macro forms. On the gravity side it goes the other way. We have
the macro theory, Einstein's general relativity. In our view
general relativity is the hydrodynamic regime of the fundamental
theory, with the metric and the connection forms as the
collective variables in this long wavelength regime, which are
likely to lose their meaning and usefulness as we probe into much
shorter scales. We want the microscopic theory of spacetime
structure and dynamics. \footnote{This theory is usually referred
to as quantum gravity, but it does not mean that quantizing the
metric or the connection functions will lead to this
micro-theory. In fact if these were the hydrodynamic collective
variables doing so will only give us a theory describing the
quantized modes of hydrodynamic excitations, not about the
microscopic structure of spacetimes \cite{grhydro}. (A similar
viewpoint is expressed by Jacobson \cite{Jac} from a different
angle.)}

Our strategy is to look closely into the quantum and statistical
mechanical features of the matter field in deepening levels and
see what this implies on the spacetime structure at the
corresponding levels. (This is different from the induced gravity
program \cite{indgra} although the spirit is similar). \footnote{
I made this choice two decades ago because the geometric
structure (left hand side) being made of marble, according to
Einstein, is beautiful to look at but difficult to chisel into,
whereas the sandstone structure (right hand side) of matter,
though not as elegant, is more malleable and easier to work
with.} Thus we work on the quantum matter field from both ends of
its structure, the micro structure described by quantum field
theory and the macro structure described by hydrodynamics. In
statistical physics it is well-known that intermediate regimes
exist between the long-wavelength hydrodynamics limit and the
microdynamics
\footnote{They are usually lumped together and called the kinetic
regime, but I think there must be distinct kinetic collective
variables depicting recognizable metastable intermediate
structures in this vast interim regime (see \cite{Spohn} for a
more fine-grained description)}. Following this pathway we need
to perform \textbf{three tasks}: First, understand the
statistical mechanics of interacting quantum fields in the matter
sector. Second, find out from statistical mechanical
considerations if there is a corresponding ordering of structure
in the gravity sector, starting with Einstein's general relativity
not just as geometro-dynamics \cite{GMD} but as
geometro-hydrodynamics \cite{grhydro}. Third, construct such a
structure for gravity leading to the micro theory of spacetime. A
figurative conception is that we are given two knotted vertical
poles representing spacetime and matter to build a ladder, with
each rung connecting the corresponding knots on two sides
representing a distinct level of structure. The lowest rung
appearing to us, creatures living at low energy, is
hydrodynamics; the next rung is semiclassical gravity, the third,
stochastic gravity. Stochastic gravity actually entails all the
higher rungs between semiclassical and quantum gravity, much like
the BBGKY or the Dyson-Schwinger hierarchy representing kinetic
theory of matter fields. One should then ask:  What are the
salient features of this kinetic theory regime of spacetime
structure? What can it reveal about the microscopic theory of
spacetime structure?  The threads sustaining both the vertical
and horizontal structures in this conceptual ladder are
\textbf{two basic issues}: micro/macro interface and
quantum/classical correspondence. That is why mesoscopic physics
also enters into our consideration \cite{meso} in a central way.

Let me indicate three signals flying across these two vertical
poles of the ladder which inspired us to the construction of the
stochastic gravity theory. 1) \textbf{Dissipation}: Our work in
the 80's on backreaction of particle creation from quantum matter
fields in cosmological spacetimes or background fields showed the
appearance of dissipation in the background spacetime or field
dynamics \cite{CH87,CH89,CV94}. 2) \textbf{Noise and
Fluctuation}: In trying to understand the statistical mechanical
meaning of this dissipation we were led to the understanding that
there should be a noise term arising from the fluctuations of the
quantum matter field \cite{Physica} and correspondingly there
should be an equation capturing the effect of noise on metric
fluctuations: the Einstein-Langevin equation \cite{ELE} which
forms the centerpiece of stochastic gravity
\cite{stogra,HVErice,MV1}. 3) \textbf{Decoherence and
Stochasticity}: One important understanding which came out of
research on quantum to classical transition (which began in the
late 80's for us working on quantum cosmology) was to recognize
that a new stochastic regime necessarily lie between the
semiclassical and the quantum in almost all physical processes.
Noise is instrumental to decoherence in the transition from
quantum to classical, resulting in a classical stochastic
dynamics.\cite{GelHar2,Tsukuba,HPZ1}

The question we wish to focus on here is how to relate the
stochastic regime to the quantum regime, first for matter field
and then for spacetime geometry. We first give a sketch of the
two poles here: stochastic gravity and kinetic field theory.

\subsection{From Semiclassical to Stochastic Gravity}

In semiclassical  gravity  the classical spacetime (with metric
$g_{ab}$) is driven by  the expectation value $\langle \rangle$
of the stress energy  tensor $T_{ab}$  of a quantum field with
respect to some quantum state.  One  main task in the 70's was to
obtain a regularized expression  for this quantum source of  the
semiclassical Einstein equation (SCE)
In stochastic gravity of the 90's 
the additional effect of fluctuations of the stress energy tensor
\cite{Ford82,KuoFor,PH97,HP012} enters which induces metric
fluctuations in the classical spacetime described by the
Einstein-Langevin equation  (ELE) \cite{ELE}. This stochastic
term measures the fluctuations of quantum sources (e.g., arising
from the difference of particles created in neighboring histories)
and  is intrinsically linked to the dissipation in the dynamics
of spacetime by a fluctuation -dissipation relation 
which embodies the full backreaction effects
of quantum fields on classical spacetime.

The stochastic semiclassical Einstein equation, or
Einstein-Langevin equation, takes on the form
\begin{eqnarray}
     G_{ab}[g] + \Lambda g_{ab}
        &=& 8\pi G ( {T_{ab}}^c +  {T_{ab}}^{qs})
           \nonumber \\
        T_{ab}^{qs}
        &\equiv& \langle T_{ab} \rangle_{q} +  T_{ab}^{s}
   \label{ELE}
\end{eqnarray}
where $G_{ab}$ is the Einstein tensor associated with $g_{ab}$
and $\Lambda, G$ are the cosmological and Newton constants
respectively. Here we use the superscripts c, s, q to denote
classical, stochastic and quantum respectively. The new term
$T_{ab}^{s}= 2 \tau_{ab}$ which is of classical stochastic nature
measures the  fluctuations of the energy momentum tensor of the
quantum field. To see what $\tau_{ab}$ is, first define
\begin{equation}
{\hat t}_{ab}(x) \equiv {\hat T}_{ab}(x) - \langle {\hat
T}_{ab}(x)\rangle \hat I \label{that}
\end{equation}
which is a tensor operator measuring the deviations from the mean
of the stress energy tensor in a particular state. We are
interested in the correlation of these operators at different
spacetime points. Here we focus on the stress energy
(operator-valued) bi-tensor $\hat t_{ab} (x) \hat t_{c'd'}(y)$
defined at nearby points $(x, y)$. A bi-tensor is a geometric
object that has support at two separate spacetime points. In
particular, it is a rank two tensor  in the tangent space at $x$
(with unprimed indices) and in the tangent space at $y$ (with
primed indices).

The noise kernel $N_{abc'd'}$ bitensor is defined as
\begin{equation}
4 N_{abc'd'} (x,y) \equiv {1\over 2} \langle \{ {\hat t}_{ab}(x),
{\hat t}_{c'd'}(y) \} \rangle \label{D4}
\end{equation}
where $\{ \}$ means taking the symmetric product.   In the
influence functional (IF) \cite{if} or closed-time-path (CTP)
effective action approach \cite{ctp} the noise kernel appears in
the real part of the influence action \cite{Banff}.The noise
kernel defines a real classical Gaussian stochastic symmetric
tensor field $\tau_{ab}$ which is characterized to lowest order
by the following relations,
\begin{equation}
\langle\tau_{ab}(x)\rangle_s=0,\ \ \ \ \langle \tau_{ab}(x)
\tau_{c'd'}(y)\rangle_s= N_{abc'd'}(x,y), \label{D6}
\end{equation}
where $\langle\,\rangle_s$ means taking a statistical average
with respect to the noise  distribution $\tau$ (for simplicity we
don't consider higher order correlations).  Since $\hat T_{ab}$
is self-adjoint, one can see that $N_{abc'd'}$ is  symmetric,
real, positive and semi-definite.  Furthermore, as a consequence
of (\ref{D4}) and the conservation law $\nabla^a \hat T_{ab}=0$,
this stochastic tensor $\tau_{ab}$ is divergenceless in the sense
that $\nabla^a \tau_{ab}=0$ is a deterministic zero field. Also
 $g^{ab}{\tau}_{ab}(x) = 0$,  signifying that there is no
stochastic correction to the trace anomaly for massless conformal
fields where $T_{ab}$ is traceless. (See \cite{MV1}). Here all
covariant derivatives are taken with respect to the background
metric $g_{ab}$ which is a solution of the semiclassical
equations. Taking the statistical average of (\ref{ELE}) with
respect to the noise distribution $\tau$, as a consequence of the
noise correlation relation (\ref{D6}),
\begin{equation}
          \langle T_{ab}^{qs}
          \rangle_s
        = \langle T_{ab} \rangle_q
\label{meant}
\end{equation}
we recover the semiclassical Einstein equation 
which is (\ref{ELE}) without the $T_{ab}^s$  term. It is in this
sense that we view semiclassical gravity as a mean field theory.

\subsection{Kinetic Theory of Interacting Quantum Fields}
\label{IB}

Our viewpoint here is motivated by classical kinetic theory. In
the dynamics of a dilute gas \cite{Akhiezer,Balescu,Spohn} the
exact Newton's or Hamilton's equations for the evolution of a
many body system may be represented by a Liouville equation for
the total distribution function or the BBGKY hierarchy for the
partial (n- particle) distributions. This reformulation is only
formal, which involves no loss of information or predictability.
Physical description of the dynamics comes from truncating the
BBGKY hierarchy and introducing a causal factorization condition
(`slaving')  whereby the higher order correlations are
substituted by functionals of the lower order ones with some
causal boundary conditions (like the molecular chaos assumption)
ingrained, which accounts for the appearance of dissipation and
an arrow of time in the Boltzmann equation.

For illustrative purpose here we can represent quantum gravity  as
an interacting quantum field (of fermions?) and we shall traverse
this passage using the \textbf{correlation dynamics} from the
\textbf{master effective action}.  The master effective action is
a functional of the whole string of Green functions of a field
theory whose variation generates the Schwinger -Dyson hierarchy.
There are \textbf{two aspects} in this problem: coherence of a
field as measured by its correlation (for quantum as well as
classical), and quantum to classical transition. We wish to treat
both aspects with a quantum version of the correlation (BBGKY)
hierarchy, the Schwinger-Dyson equations. There are \textbf{three
steps} involved: First, show how to derive the kinetic equations
from quantum field theory -- or to go from Dyson to Boltzmann
\cite{CH88}. Second, show how to introduce the open system
concept to the hierarchy. For this we need to introduce the
notion of `slaving' in the hierarchy, which renders a subset made
up of a finite number of lower order correlation functions as an
effectively open system,  where it interacts with the environment
made up of the higher correlation functions. Third, show why
there should be a stochastic term in the Boltzmann equation when
influence of the higher correlation functions are included.


The first step was taken in the mid-80's,  when, amongst many
authors  (see \cite{kft80} for earlier work and \cite{kft90} for
recent developments) Calzetta and I \cite{CH88}, showed how the
quantum Boltzmann equation arises as a description of the
dynamics of quasiparticles in the kinetic limit of quantum field
theory. The main element in the description of a nonequilibrium
quantum field is its Green functions, whose dynamics is given by
the Dyson equations.
 For the second step, we showed  \cite{cddn} how the
coarse-grained (truncation with slaving) n-point correlation
functions behave like an effectively open system. It is easy to
illustrate this idea with the lowest order elements in the
Schwinger-Dyson  hierarchy of correlation functions, consisting
of the mean field and the two point function. They are deducible
from the CTP (closed-time-path) 2PI (two-particle-irreducible)
effective action which has been used to derive the Boltzmann
equation \cite{CH88} in kinetic field theory, for problems in
critical dynamics and many nonequilibrium quantum field
processes.  Generalizing from $n=2$ to $\infty$ yields the master
effective action (MEA). In general the full MEA is required to
recover full (including phase) information in a quantum field. If
we now view the problem in this framework we can see how
dissipation and fluctuations arise when the hierarchy is
truncated and the higher correlations are slaved (we refer to
these two procedures combined as coarse-graining), in the same
way how Boltzmann equation is derived from the BBGKY hierarchy.
In \cite{cddn}  we 1) gave a formal construction of the master
effective action, 2) showed how truncation in nPI is related to
loop expansion and 3) how `slaving' leads to dissipation.

Now for the third step: our assertion is that there should also be
a noise term present as source in addition to the collision term
in the Boltzmann equation, making it a stochastic Boltzmann, or
Boltzmann-Langevin (BL) equation. Can we find the noise in the
correlation hierarchy? Note that we are now following the
Boltzmann paradigm of effectively open systems, not the Langevin
paradigm of stipulated open systems. It is more difficult to find
the noise from the BBGKY hierarchy (an example for classical gas
is that of Kac and Logan \cite{KacLogan}) or the Schwinger-Dyson
series of correlation functions of interacting quantum fields
than finding noise in a well-defined environment (e.g., \textit{a
la} Feynman-Vernon). We  were partially successful in identifying
such a correlation noise \cite{stobol} arising from the slaving of
the higher correlation functions and proving a
fluctuation-dissipation relation for these correlation noises.
With the BL equation one can use the correlation hierarchy to
infer the quantum microdynamics. From the self-consistency
between the matter (quantum field) and the spacetime as suggested
by Einstein's equation one can construct a parallel hierarchy of
metric correlation functions induced by the matter field. From
this hierarchy, beginning with stochastic gravity, one could
begin to unravel the microstructure of spacetime.


This summarizes the two theories which make up the poles of our
ladder. We will now try to apply the statistical mechanical ideas
to find a route from stochastic to quantum gravity. In Sec. 2 we
give a sketch of the stochastic regime in relation to the
semiclassical and quantum regimes, with the specific aim of
seeking a pathway to quantum gravity. In Sec. 3 we highlight the
issues in the statistical mechanics of interacting quantum fields
which we need to address. In Sec. 4 we give an example of how the
correlation hierarchy of quantum fields can be applied to a
possible resolution of the black hole information loss paradox,
based on ideas proposed before.
Finally in Sec. 5 we develop further some issues anticipated en
route, i.e., the reconstruction of coherence via the correlation
hierarchy of spacetime fluctuations. This involves ideas from
kinetic field theory and mesoscopic transport theory. We end with
a task summary.

\section{Stochastic in relation to Semiclassical and Quantum
Gravity}

We see that stochastic semiclassical gravity provides a relation
between noise in quantum fields and metric fluctuations. While the
semiclassical regime describes the effect of a quantum matter
field only through its mean value (e.g., vacuum expectation
value), the stochastic regime includes the effect of fluctuations
and correlations.
 We believe that precious new information resides in the two-point functions of the
stress energy tensor which may reflect the finer structure of
spacetime at a scale when information provided by its mean value
as source (semiclassical gravity) is no longer adequate. To
appreciate this, it is perhaps instructive to examine the
distinction among these three  theories: stochastic gravity in
relation to semiclassical and quantum  gravity \cite{MV1,stogra}.
The following observation (adopted from \cite{stogra}) will also
bring out two other related concepts of correlation (in the
quantum field) and coherence (in quantum gravity).

\subsection{Classical, Stochastic and Quantum}

For concreteness we consider the example of gravitational
perturbations $h_{ab}$ in a  background spacetime with metric
$g_{ab}$ driven by the expectation value of the energy momentum
tensor of a scalar field $\Phi$, as well as its fluctuations
${\hat t}_{ab}(x)$. Let us compare the stochastic with the
semiclassical and quantum equations  of motion for the metric
perturbation (weak but deterministic) field $h$. (This schematic
representation was made by E. Verdaguer in Peyresq 3). The
semiclassical equation is given by
\begin{equation}
\Box h =16\pi G \langle \hat T \rangle
\end{equation}
where $\langle \rangle$ denotes taking the quantum average (e.g.,
the vacuum expectation value)  of the operator enclosed. Its
solution can be written in the form
\begin{equation}
h = \int C  \langle \hat T\rangle, ~~~~~ h_1h_2 = \int\int C _1C
_2 \langle \hat T\rangle \langle\hat T\rangle.
\end{equation}
The quantum (Heisenberg) equation
\begin{equation}
\Box \hat h =16 \pi G  \hat T
\end{equation}
has solutions
\begin{equation}
\hat h = \int C  \hat T, ~~~~~ \langle \hat h_1 \hat h_2\rangle =
\int \int C _1 C _2 \langle \hat T \hat
    T \rangle_{\hat h, \hat \phi}
\end{equation}
where the average is taken  with respect to the quantum
fluctuations of both the gravitational ($ \hat g $) and the
matter ($\hat \phi$) fields. Now for the stochastic equation, we
have
\begin{equation}
\Box h =16 \pi G ( \langle \hat T \rangle + \tau)
\end{equation}
with solutions \footnote{In this schematic form we have not
displayed the homogeneous
 solution carrying the information of the (maybe random) initial condition.
This solution will  exist in general, and may even be dominant if
dissipation is weak. When both the uncertainty in initial
conditions and the stochastic noise are taken into account, the
Einstein - Langevin formalism reproduces the exact graviton two
point function, in the linearized approximation. Of course, it
fails to reproduce the expectation value of observables which
could not be written in terms of graviton occupation numbers, and
it is in these aspects which one needs to reconstruct to get to
full quantum gravity. This comment was made by E. Calzetta in a
correspondence to the author.}

\begin{equation}
h = \int C  \langle \hat T \rangle + \int C  \tau, ~~~~~ h_1 h_2
= \int\int C _1 C _2 [ \langle \hat T \rangle \langle \hat T
\rangle + (\langle \hat T \rangle \tau + \tau \langle \hat T
\rangle) + \tau \tau]
\end{equation}
Now take the noise average $\langle \rangle_\tau$ . Recall that
the noise is defined in terms of the stochastic sources $\tau$ as
\begin{equation}
\langle \tau \rangle_\tau = 0, ~~~~ \langle
\tau_1\tau_2\rangle_\tau \equiv
    \langle \hat T_1 \hat T_2 \rangle - \langle \hat T_1\rangle
\langle \hat T_2\rangle
\end{equation}
we get
\begin{equation}
\langle h_1 h_2 \rangle_\tau = \int \int C _1 C _2 \langle \hat T
\hat T \rangle_{\hat \phi}
\end{equation}
Note that the correlation of the energy momentum tensor appears
just like in the quantum case, but the average here is  over
noise from quantum fluctuations of the matter field alone.

\subsection{Noise and Fluctuations, Correlations and Coherence}

Comparing the equations above depicting the semiclassical,
stochastic and quantum regimes, we see first that in the
semiclassical case, the classical metric correlations is given by
the product of the vacuum expectation value of the energy
momentum tensor whereas in the quantum case it is given by the
quantum average of the correlation of metric (operators) with
respect to the fluctuations in both the matter and the
gravitational fields. In the stochastic case the form is  closer
to the quantum case except that the quantum average is replaced
by the noise average, and the average of the energy momentum
tensor is taken with respect only to the matter field. The
important improvement over semiclassical gravity is that it now
carries information on the correlation of the energy momentum
tensor of the fields and its induced metric fluctuations. Thus
stochastic gravity contains information about the correlation of
fields (and the related phase information) which is absent in
semiclassical gravity. Here we have invoked the relation between
{\it fluctuations} and {\it correlations}, a variant form of the
fluctuation-dissipation relation. This feature moves stochastic
gravity  closer than semiclassical gravity to quantum gravity in
that the correlation in quantum field and geometry fully present
in quantum gravity is partially retained in stochastic gravity,
and the background geometry has a way to sense the correlation of
the quantum fields through the noise term in the
Einstein-Langevin equation, which shows up as metric fluctuations.

By now we can see that `noise' as used in this more precise
language and broader context is not something one can arbitrarily
assign  or relegate, as is often done in ordinary discussions,
but it has taken on a deeper meaning in that it embodies the
contributions of the higher correlation functions in the quantum
field. It holds the key to probing the quantum nature of spacetime
in this vein. We begin our studies here with the lowest order
term, i.e., the 2 point function of the energy momentum tensor
which contains the 4th order correlation of the quantum field (or
gravitons when they are considered as matter
source).\footnote{Although the Feynman- Vernon way can only
accomodate Gaussian noise of the matter fields and takes a simple
form for linear coupling to the background spacetime, the notion
of noise can be made more general and precise. For an example of
more complex noise associated with more involved backreactions
arising from strong or nonlocal couplings, see Johnson and Hu
\cite{JH1}} Progress is made now on how to characterize the higher
order correlation functions of an interacting field
systematically from the Schwinger-Dyson equations in terms of
`correlation noise' \cite{cddn,stobol}, after the BBGKY hierarchy.
This will prove to be useful for a correlation dynamics
/stochastic semiclassical approach to quantum gravity
\cite{stogra}.

Thus noise carries information about the correlations of the
quantum field. One can further link {\it correlation} in quantum
fields to {\it coherence} in quantum gravity. This stems from the
self-consistency required in the backreaction equations for the
matter and spacetime sectors. The Einstein-Langevin equation is
only a partial (low energy) representation of the complete theory
of quantum gravity and fields. There, the coherence in the
geometry is related to the coherence in the matter field, as the
complete quantum description should be given by a coherent wave
function of the combined matter and gravity sectors.
Semiclassical gravity forsakes all the coherence in the quantum
gravity sector. Stochastic gravity captures only partial
coherence in  the quantum gravity sector via the correlations in
the quantum fields. Since the degree of coherence can be measured
in terms of correlations, our strategy for the semiclassical
stochastic gravity program  is to unravel the higher correlations
of the matter field, going up the hierarchy starting with the
variance of the stress energy tensor,  and through its linkage
with gravity (the lowest rung provided by the Einstein equation),
retrieve whatever quantum attributes (partial coherence) of
gravity left over from the high energy behavior above the Planck
scale.  Thus in this approach, focusing on the noise kernel and
the stress energy tensor two point function is our first step
beyond mean field (semiclassical gravity) theory towards probing
the full theory of quantum gravity.

In recent years there has been increasing attention paid to the
effects of a fluctuating spacetime such as related to structure
formation \cite{strfor}, black hole event horizons and Hawking
radiation
\cite{CanSci,Mottola,HRS,CamHu,HuShio,Ford,Sorkin,BFP,MasPar,Par},
and spacetime foam \cite{STfoam}. Only a small subset of work has
included backreaction considerations, such as that of induced
metric fluctuations from a quantum scalar in Minkowski spacetime
treated by \cite{MV2}, which is closest to the spirit of our
description above. Better understanding in the relation of
quantum correlation functions with its classical \cite{CAcorf} and
stochastic \cite{CRV} counterparts is also conducive to the
pursuit of our program.

To continue this line of thought to seek a pathway to the
micro-structure of spacetime (our definition of quantum gravity)
via the correlation hierarchy, we need some more discussions on
the statistical mechanics of quantum fields.

\section{Statistical Mechanics of an Interacting Quantum Field}

Though interacting quantum fields is a familiar subject which one
learns in the first lessons of quantum field theory, not much has
been explored in its statistical mechanical content. It is
surprisingly rich, much like the role an ordinary box of gas
molecules plays in Boltzmann's sophisticated theory of kinetics,
dissipation and arrow of time. In Sec. \ref{IB}  we mentioned the
three steps taken to obtain a stochastic Boltzmann equation and
the two aspects: The micro-macro relation and the quantum to
classical transition. Here we expand on those two aspects.
Details can be found in  \cite{CH88,dch,cddn,stobol}


 The statistical mechanical properties of interacting quantum
fields can be studied in terms of the {\it dynamics of the
correlation functions}. The full dynamics of an interacting
quantum field may be described by means of the Dyson- Schwinger
equations governing the infinite hierarchy of Wightman functions
which measure the correlations of the field. This hierarchy of
equations can be obtained from the variation of the infinite
particle irreducible, or {\it `master' effective action} (MEA).
Truncation of this hierarchy gives rise to a quantum subdynamics
governing a finite number of  correlation functions (which
constitute the `system'). `Slaving' refers to expressing the
higher order correlation functions (which constitute the
`environment') in terms of the lower-order ones by functional
relations such as the `factorization' and the imposition of causal
conditions (`molecular chaos' assumption). The latter condition
induces {\it dissipation} in the dynamics of the subsystem. In
addition to the collision integrals there should also be a
stochastic source representing the fluctuations of the
environment, which we call the {\it `correlation noise'}. We
posit that at each level of the hierarchy these two aspects should
be related by a fluctuation-dissipation relation.

This is the quantum field equivalent of the BBGKY hierarchy in
Boltzmann's theory. Any subsystem involving a finite number of
correlation functions defines an effective theory. The relation
of loop expansion and correlation order was expounded
\cite{cddn}. We see that ordinary quantum field theory which
involves only the mean field and a two-point function, or any
finite-loop effective action in a perturbative theory are, by
nature, also effective theories. Histories defined by lower-order
correlation functions can be decohered by the noises from the
higher order functions and acquire quasi-classical stochastic
attributes.  We think this scheme invoking the correlation order
is a natural way to describe the quantum to classical transition
for a closed system as it avoids {\it ad hoc} stipulation of the
system-environment split. It is in this spirit that our kinetic
theory approach to quantum gravity works.

\subsection{ Statistical Mechanics Aspect: Correlation Dynamics
in the BBGKY or Dyson-Schwinger Hierarchy}

So we want to describe an interacting quantum field in terms of
the (infinite number of) correlation functions. In the consistent
history approach \cite{conhis} to quantum mechanics, we view the
`mean' field not as the actual expectation value of the field,
but rather as representing the local value of the field within
one particular history. Quantum evolution encompasses the
coherent superposition of all possible histories  and these
quantities are subject to fluctuations. This naturally introduces
into the theory stochastic elements, which has hitherto been
largely ignored in the usual description of quantum field theory.
The usual loop expansion in quantum field theory is replaced in
our scheme by the correlation functions up to a certain order
acting as independent variables, along with the `mean' field,
which themselves are subject to fluctuations.

Our starting point is the well-known fact that the set of all
Wightman functions (time ordered products of field operators)
determines completely the quantum state of a field \cite{Haag}.
Instead of following the evolution of the field in any of the
conventional representations (Schr\"odinger, Heisenberg or
Dirac's), we focus on the dynamics of the full hierarchy of
Wightman functions. To this end it is convenient to adopt
Schwinger's ``closed time-path'' techniques \cite{ctp}, and
consider time ordered Green functions as a subset of all Green
functions path- ordered along a closed time loop. The dynamics of
this larger set is described by the Dyson - Schwinger equations.

We first showed that the Dyson- Schwinger hierarchy may be
obtained via the variational principle from a functional which we
call the `Master Effective Action' (MEA). This is a formal action
functional where each Wightman function enters as an independent
variable. We then showed that any field theory based on a finite
number of (mean field plus ) correlation functions can be viewed
as a subdynamics of the Dyson- Schwinger hierarchy. The
specification of a subdynamics involves two steps: First, the
hierarchy is {\it truncated} at a certain order. A finite set of
variables,  say, the lowest nth order correlation functions, is
identified to be the `relevant' \cite{projop} variables, which
constitute the subsystem. Second, the remaining `irrelevant' or
`environment' variables, say, the n+1 to $\infty$ order
correlation functions, are {\it slaved} to the former. Slaving
(imposition of the factorization and causal conditions) means
that irrelevant variables are substituted by set functionals of
the relevant variables.  The process of extraction of a
subdynamics from the Dyson- Schwinger hierarchy shows up at the
level of the effective action, where the MEA is truncated to a
functional of a finite number of variables. The finite effective
actions so obtained (the influence action \cite{if}) are
generally nonlocal and complex, which is what gives rise to the
noise and dissipation in the subdynamics. Moreover, since the
slaving process generally involves a choice of causal initial
conditions, an arrow of time (irreversibility) appears in the
cloak of dissipation in the subdynamics \cite{projop}.

\subsection{Quantum-Classical Aspect: Decoherence of Correlation
History}

Decoherence is brought on by the effect of a coarse-grained
environment (or `irrelevant' sector) on the system (or the
`relevant' sector). In the open systems philosophy, this split is
imposed by hand, as when some of the fields, or the field values
within a certain region of spacetime, are chosen as relevant.
This is not appropriate for treating close or nearly close
systems which is what we have at hand. Instead, we adopt the
consistent histories view \cite{conhis} of the quantum to
classical transition problem, but instead of introducing
projections along histories as in the decoherent histories scheme
\cite{dechis} \footnote{A closely related ongoing program is
deducing hydrodynamic variables and dynamics \cite{BHLM,CAhydro} .
It would be interesting to ask what the kinetic regime would be
like in the decoherent histories scheme.} we let the correlation
ordering in the full hierarchy act as its own projection. For any
given degree of accuracy of observation or measurement which can
be carried out on the close system there is a corresponding
correlation order which is commensurate with it. What results is
an effectively open system described by an effective theory
characteristic of the physics of the energy or length scale
defined by the highest correlation order effective in the system.
There is no need to select {\it a priori} a relevant sector
within the theory. This is the main philosophy behind the
decoherence of correlation histories (DCH) \cite{dch} approach.
In this framework, decoherence occurs as a consequence of the
fluctuations in the higher order correlations and results in a
classical dissipative dynamics of the lower order correlations.

Applying the DCH scheme to interacting quantum fields we consider
the full evolution of the field described by the Dyson- Schwinger
hierarchy as a fine- grained history while histories where only a
finite number of Wightman functions are freely specified (with
all others slaved to them) are  coarse-grained. We have shown
that the finite effective actions obtained for the subsystems of
lower-order correlations are related to the decoherence
functional between two such histories of correlations \cite{dch},
its acquiring an imaginary part signifies the existence of noise
which facilitates decoherence. Thus decoherence of correlation
histories is a necessary condition for the relevance of the
classical theory as a description of observable phenomena. It can
be seen that if the classical theory which emerges from the
quantum subdynamics  is dissipative, then it must also be
stochastic. \footnote{ Because the fundamental variables are
quantum in nature, and therefore subject to fluctuations, a
classical, dissipative dynamics would demand the accompaniment of
stochastic sources in agreement with the `fluctuation -
dissipation theorems', for, otherwise, the theory would permit
unphysical phenomena as the damping away of zero - point
fluctuations. } From our correlation history viewpoint, the
stochasticity is in fact not confined to the field
distributions-- the correlation functions would become stochastic
as well \cite{KacLogan}.

Taking the truncated hierarchy as an effectively open system we
can relate the imaginary part of the finite effective actions
describing the truncated correlations to the auto-correlation of
the stochastic sources, i.e., correlation noises.  From the
properties of the complete (unitary) field theory which
constitutes the closed (untruncated) system, one can show that
the imaginary part of the effective action is related to the
nonlocal part of the real part of the effective action which
depicts dissipation. This is the origin of the fluctuation-
dissipation relations for non-equilibrium systems \cite{fdrsc}.
With this additional stochastic source which  drives the
classical fields and their correlation functions, we obtain a set
of Boltzmann-Langevin equations.

We thus see the interplay of two major paradigms in nonequilibrium
statistical mechanics: the Boltzmann-BBGKY and the
Langevin-Fokker-Planck descriptions and the intimate connection
between dissipation, fluctuations/noise and decoherence
\cite{Tsukuba,GelHar2}, now manifesting in the hierarchy of
correlations which defines the graded subsystems.





\section{Correlations in a Quantum Field and Black Hole
Information Paradox}

We now give an example of how the idea of correlation hierarchy
can be applied to understand certain puzzling phenomena in
semiclassical gravity, such as the black hole information loss
paradox. The use of kinetic field theory ideas was first proposed
in \cite{BHinfo}. Similar viewpoint can be found in
\cite{Wilczek,AngRecoh}.

We assume that the black hole and the quantum field with Hawking
radiation together constitute a closed system. Even though the
quantum field might be assumed to be free in the beginning,
interaction still exists in its coupling with the black hole,
especially when strong backreaction is included. We can model
this complete system by an interacting quantum field. A particle-
field system is a particular case of it. Of course a black hole
is different from a particle. In this modeling, we will first
explain how information is registered or `lost' in an interacting
quantum field, then the distinct features of a black hole and
finally, the information loss paradox of black hole systems.

To approach the black hole information loss paradox we need to
understand three conceptual points: \footnote{The idea introduced
in \cite{BHinfo} which is reproduced here was based on several
components developed in varying depths since 1986: The development
of the correlation dynamics in quantum fields formalism was done
with Esteban Calzetta \cite{cddn,stobol}, that of viewing black
hole radiance and inflation as exponential scaling
\cite{HuEdmonton,Dalian} was explored with Yuhong Zhang
\cite{cgea,jr,CHM} partly based on work on critical phenomena
done earlier with Denjoe O'Connor \cite{HuO'C}.  For general
background on this issue, see the review of Page \cite{Page}
which contains a comprehensive list of references till 1993, and
Bekenstein \cite{Bekenstein}
more recently. }
\\
1) How does one characterize the information content of a quantum
field (interacting, as a model for the black hole -
quantum field closed system, with backreaction)\\
2) How does the information flow from one part of this closed
system (hole) to another (field) and vice versa in the lifetime
of the black hole? Does information really get `lost'? If yes,
where has it gone?
Can it be retrieved? If no, where does it reside?\\
3) What is special about black hole radiation system as distinct
from ordinary particle / field system?

\subsection{Correlation functions as registrar  and
correlation dynamics as flow-meter of information in quantum
fields}

The set of correlation functions provides us with the means to
register the information content of a quantum field. As mentioned
above, the complete set of $ n = \infty $ correlation (Wightman)
functions carries the complete information about  the quantum
state of the field. A subset of it which defines the subsystem,
such as the mean field and the 2-point function, as is used in
the ordinary description of (effective) field theory,  carries
only partial information. The missing information resides in the
correlation noise, and manifests as dissipation in the subsystem
dynamics. In this framework, the entropy of an incompletely
determined quantum system is simply given by $ S = - Tr
\rho_{red} ln \rho _{red}$, where the reduced density matrix of
the subsystem (say,  consisting of the lower correlation orders)
is formed by integrating out the environmental variables (the
higher correlation orders) after the hierarchy is truncated with
causal factorization conditions.

While the set of correlation functions act as a registrar of
information of the quantum system, keeping track of how much
information resides in what order, the dynamics of correlations
as depicted by the hierarchy of equations of motion derived from
the master effective action depicts the flow of information from
one order to another, up or down or criss-crossing the hierarchy.
Correlation dynamics has been proposed for the description of
many body systems before \cite{Balescu}, and applied to molecular
and plasma kinetics.  In the light of the above theoretical
description we see this scheme as a potentially powerful way to
systemize quantum information, i.e., keeping track of the content
and flow of information in a coherent or partially coherent
quantum system.

\subsection{Information appears lost to subsystems of lower order
correlations -- `missing' information stored in higher order
correlations}

Most measurements of a quantum field system are of a local or
quasilocal nature. If one counts the information content of a
system based on the mean field and the lowest order correlation
functions, as in the conventional way (of defining quantum field
theory in terms of, e.g., 2PI effective action),  one would miss
out a good portion of the information in the complete system, as
much of that now resides in the higher order correlation
functions in the hierarchy. These invoke nonlocal properties of
the field, which are not easily accessible in the ordinary range
of accuracy in measurements. Such an observer would then report a
loss of information in his way of accounting (which is taken to
be in agreement with other observers with the same level of
accuracy of measurement). Only observers which have access to all
orders (the `master' in the master effective action) would be
able to see the complete development of the system and be able to
tell when the nth order observer begins to lose track of the
information count and report an information loss. This is more
easily seen in molecular dynamics: For observers confined to
measuring one particle distribution functions (truncation of the
BBGKY hierarchy) and with the molecular chaos assumptions
implicitly invoked (causal factorization condition, or `slaving'),
he would report on information loss (not if it is a simple
truncation with no slaving, as the subsystem will then be
closed). This is how Boltzmann reasons out the appearance of
dissipation in ordinary macroscopic physical phenomena. The same
can be said about measurement of quantum systems.

\subsection{Exponential scaling in Hawking effect facilitates information
transfer to the higher correlations. Black hole with its radiation
contains full information, but retrieval requires probing the
higher order nonlocal properties of the field}

How is this scheme useful in addressing the black hole information
problem?  How is the black hole / quantum field system different
from the ordinary cases? The above scheme can explain the
apparent loss of information in a quantum system, but there is an
aspect distinct to black holes or systems emitting thermal
radiance. It was observed that all mechanisms of emission of
(coherent) thermal radiance such as the Hawking effect in black
holes, or the Unruh effect in accelerated detectors, involve an
exponential redshifting process in the system
\cite{HuEdmonton,Dalian}. This can be compared to the scaling
transformation in treating critical phenomena \cite{cgea,jr,CHM}.
After sufficient exponential redshifting (at late times of
collapse) and the black  hole is emitting thermal radiance, the
system has reached a state equivalent to the approach to a
critical point in phase transition. There, the physical
properties of the system are dominated by the infrared behavior,
and as such, the lowest order correlation functions are no longer
sufficient to characterize the critical phenomena. The
contribution of higher order correlation functions would become
important.
Note that in ordinary situations, only the mean field and the 2
or 3  point correlation functions are needed to give an adequate
description of the dynamics of the system. But for black holes or
similar systems where exponential red-shifting is at work, higher
order correlations are readily activated. The information content
profile for a quantum field in the presence of a black hole would
be very different from ordinary systems, in that it is more
heavily populated in the higher order correlations. If one
carries out measurements which are only sensitive to the lower
correlations, one would erroneously conclude that there is
information loss.

So, following the correlation dynamics of the black hole / field
system, while the state of the combined system remains the same
as it had begun, there is a continuous shifting of information
content from the black hole to the higher correlations in the
field as it evolves. Correlation dynamics of fields can be used
to keep track of this information flow. We speculate that the
information content of the field will be seen to shift very
rapidly from low correlation orders to the higher ones as Hawking
radiation begins and continues. The end state of the system would
have a black hole evaporated, and its information content
transferred to the quantum field, with a significant portion of
it residing in the higher order nonlocal correlations.

This is, however, not the end of the story for the correlation
dynamics and information flow in the field. The information
contained in the field will continue, as it does in general
situations, to shift across the hierarchy. As we know from the
BBGKY description of molecular dynamics, after  the higher
correlation orders in the hierarchy have been populated -- and
for systems subjected to exponential red-shifting this condition
could be reached relatively quickly -- the information will begin
to trickle downwards in the hierarchy, though far slower than the
other direction initially. The time it takes (with many
criss-crossing) for the information to return to the original
condition is the Poincare recurrence time. This time we suspect
is the upper bound for the recoherence time, the time for a
coherent quantum system interacting with some environment to
regain its coherence  \cite{AngRecoh}. It would be interesting to
work out the information flow using the correlation dynamics
scheme for a few sample systems, both classical and quantum,  so
as to distinguish the competing effects of different
characteristic processes in these systems, some quantum, some
statistical (e.g., decoherence time, relaxation time, recoherence
time and recurrence time). \footnote{Our depiction above uses the
interacting field model. Simpler cases might show somewhat
degenerate behavior. An example is the interesting result of
recoherence reported by Anglin et al \cite{AngRecoh}. We think
their reported result of a recoherence time of the order of the
relaxation time is special to the simple model of particle
free-field interaction. As the field modes couple only through
their interaction with the particle, and not amongst themselves,
there is no structure or dynamics of the information content of
the field itself, and the only time scale for it to return is via
interaction with the particle, which is why  the recoherence time
is related to the relaxation time of the particle.  We expect in
more general and complex systems (thus excluding many spin
systems) the recoherence time is much longer than the relaxation
time, more in the order of the recurrence time.}


\section{From Stochastic to Quantum Gravity via Metric Correlation Hierarchy}


After this expose of the notion of correlation dynamics (kinetic
theory) representation of quantum field theory, and an example of
its application to dealing with some conceptual issues, we now
return to the main progression of ideas: What then? Where is
quantum gravity?

\subsection{Quantum Coherence from Correlations of Induced Metric Fluctuations}

Let us take another look at the equations in the Section where we
compare the relation of semiclassical and the stochastic regimes
with the quantum regime. We see first that in the semiclassical
case, the classical metric correlations is given by the product
of the vacuum expectation value of the energy momentum tensor
whereas in the quantum case the quantum average of the
correlation of metric (operators) is given by the quantum average
with respect to the fluctuations in both the matter and the
gravitational fields. In the stochastic case the form is closer
to the quantum case except that now the quantum average is
replaced by the noise average, and the average of the energy
momentum tensor is taken with respect only to the matter field.
The important improvement over the semiclassical case is that it
now carries information on the correlation of the energy momentum
tensor of the fields and its induced metric fluctuations. This is
another way to see why the stochastic description  is closer to
the quantum truth. More intuitively, the difference between
quantum and semiclassical is that the latter loses all the
coherence in the quantum gravity sector. Stochastic improves on
the semiclassical situation in that partial information related
to the coherence in the gravity sector is preserved as is
reflected in the backreaction from the quantum fields and
manifests as induced metric fluctuations. That is why we need to
treat the noise terms with maximal respect. It contains quantum
information absent in the classical. The coherence in the
geometry is related to the coherence in the matter field, as the
complete quantum description should be given by a coherent wave
function of the combined matter and gravity sectors. Since the
degree of coherence can be measured in terms of correlations our
strategy is to examine the higher correlations of the matter
field, starting with the variance of the energy momentum tensor
in order to probe into or retrieve whatever partial coherence
remains in the quantum gravity sector. The noise we worked out in
the Einstein-Langevin equation above contains the 4th order
correlation of the quantum field (or gravitons when considered as
matter source) and  manifests as induced metric fluctuations.

If we view classical gravity as an effective theory, i.e., the
metric or connection functions as collective variables of some
fundamental particles which make up spacetime  in the large and
general relativity as the hydrodynamic limit, we can also ask if
there is a mid-way weighing station like kinetic theory from
molecular dynamics, from quantum micro-dynamics to classical
hydrodynamics. This transition involves both the micro to macro
transition and the quantum to classical transition, which is what
constitutes the mesoscopic regime for us.

\subsection{Stochastic Boltzmann-Einstein Equations for Spacetime Correlations and Fluctuations}

In the Introduction we have defined the master effective action
and showed its relation to the Schwinger-Dyson hierarchy. From
this one can establish a kinetic theory of nonlinear quantum
fields \cite{CH88}. Here a brief description of the difference
between an open and an effectively open system is perhaps useful.
While the Langevin equation describes an open system, its noise
source arising from coarse-graining the environment, the full
BBGKY hierarchy describes a closed system, coarse-graining
(truncation and slaving) the hierarchy yields an effectively open
system, for which Boltzmann's theory is the lowest order example.
The Boltzmann equation describes the evolution of one particle
distribution function driven by a 2 particle collision integral
with causal factorization condition (an example of slaving). Note
that truncation without slaving will not produce an open system,
the nth order correlation functions constitute a smaller closed
system. They obey an equation of the Vlasov type which is
unitary. Inclusion of a noise source representing the slaved
contributions of the higher correlation functions gives the
stochastic Boltzmann equation. The stochastic Boltzmann equation
\cite{stobol} contains features which would enable us to make
connection with the stochastic equation in semiclassical gravity.

The hierarchical structure illustrated here for interacting
quantum fields can be extended to a description of the
micro-structure of spacetime --  assuming that it can be
represented by some interacting quantum field, or its extended
version, string field theory. Starting with stochastic gravity we
can get a handle on the correlations of the underlying field of
spacetime by examining (observationally if possible, e.g.,
effects of induced spacetime fluctuations) the hierarchy of
equations, of which the Einstein-Langevin equation is the lowest
order, i.e., the relation of the mean field to the two point
function, and the two point function to the four (variance in the
energy momentum tensor), and so on. One can in principle move
higher in this hierarchy to probe the dynamics of the higher
correlations of spacetime substructure. This addresses the
correlation aspect; the quantum to classical aspect can be
treated by the decoherence of correlation histories \cite{dch}
scheme discussed in an earlier section. \footnote{In
\cite{stogra} I also brought up the relevance of the large N
expansion in gravity for comparison. There exists a relation
between correlation order and the loop order \cite{cddn}. One can
also relate it to the order in large N expansion (see, e.g.,
\cite{AarBer}). It has been shown that the leading order 1/N
expansion for an N-component quantum field yields the equivalent
of semiclassical gravity \cite{HarHor}. The leading order 1/N
approximation yields mean field dynamics of the Vlasov type which
shows Landau damping which is intrisically different from the
Boltzmann dissipation.  In contrast the equation obtained from
the nPI (with slaving) contains dissipation and fluctuations
manifestly. It would be of interest to think about the relation
between semiclassical and quantum in the light of the higher 1/N
expansions \cite{Tomb}, which is quite different from the
scenario associated with the correlation hierarchy.}

\subsection{Strongly Correlated Systems: Spacetime Conductance Fluctuations}

At this point it is perhaps useful to bring back another theme we
explored in earlier exposition of this subject matter, i.e.,
semiclassical gravity as mesoscopic physics. \footnote{ To
practitioners in condensed matter and atomic/optical physics,
mesoscopia refers to rather specific problems where, for example,
the sample size is comparable to the probing scale (nanometers),
or the interaction time is comparable to the time of measurement
(femtosecond), or that the electron wavefunction correlated over
the sample alters its transport properties, or that the
fluctuation pattern is reproducible and sample specific. Take
quantum transport. Traditional transport theory applied to
macroscopic structures are based on kinetic theory while that for
mesoscopic structures is usually based on near-equilibrium or
linear response approximations (e.g., Landauer-B\"utiker
formula). New nanodevice operations involve nonlinear,
fast-response and far-from-equilibrium processes which are
sensitive to the phases of the electronic wavefunction over the
sample size. These necessitate a new microscopic theory of
quantum transport. One serious approach is using the Keldysh
method in conjunction with Wigner functions. It is closely
related to the closed-time-path formalism we developed for
nonequilibrium quantum fields aimed for similar problems in the
early universe and black holes \cite{CH88,CHR}.}

Viewing the issues of correlations and quantum coherence in the
light of mesoscopic physics we see that what appears on the right
hand side of the Einstein-Langevin equation, the stress-energy two
point function, is analogous to conductance of electron transport
which is given by the current-current two point function. What
this means is that we are really calculating the transport
functions of the matter particles as depicted here by the quantum
fields. Following Einstein's observation that spacetime dynamics
is determined by (while also dictates) the matter (energy
density), we expect that the transport function represented by
the current correlation in the fluctuations of the matter energy
density would also have a geometric counterpart and equal
significance at a higher energy than the semiclassical gravity
scale. This is consistent with general relativity as
hydrodynamics: conductivity, viscosity and other transport
functions are hydrodynamic quantities. Here we are after the
transport functions associated with the dynamics of spacetime
structures. The Einstein tensor correlation function illustrated
in an earlier section is one such example. For many practical
purposes we don't need to know the details of the fundamental
constituents or their interactions to establish an adequate
depiction of the low or medium energy physics, but can model them
with semi-phenomenological concepts like mean free path and
collisional cross sections applicable at the lowest level (where
correlation and coherence are ignored). In the mesoscopic domain
the simplest kinetic model of transport using these concepts are
no longer accurate enough. One needs to work with
system-environment models and keep the phase information of the
collective electron wave functions. When the interaction among
the constituents gets stronger, or the probing scale gets
shorter, effects associated with the higher correlation functions
of the system begin to show up. Studies in strongly correlated
systems are revealing in these regards. For example, conductance
fluctuations obtained from the 4 point function of the current
carry important information such as the sample specific signature
and universality. Although we are not quite in a position,
technically speaking, to calculate the energy momentum 4 point
function (see, however, \cite{Shiokawa}), thinking about the
problem in this way may open up many interesting conceptual
possibilities, e.g., what does universal conductance fluctuations
mean for spacetime and its underlying constituents? (In the same
vein, I think studies of nonperturbative solutions of
gravitational wave scattering in M(atrix) theory \cite{gravscat}
or exact solutions of colliding gravitational waves (see, e.g.
\cite{DorVer,BelVer}, the Khan-Penrose or Nutku-Halil solutions,
and Misner's harmonic mapping theory \cite{harmap})  will also
reveal interesting information about the underlying structure of
spacetime beyond the hydrodynamic realm). Thus, viewed in the
light of mesoscopic physics, with stochastic gravity as a
stepping stone, we can begin to probe into the higher
correlations of quantum matter and with them the associated
excitations of the collective modes in geometro-hydrodynamics,
the kinetic theory of spacetime meso-dynamics and eventually
quantum gravity -- the theory of spacetime micro-dynamics.

\subsection{Task Summary}

To summarize, our pathway from stochastic to quantum gravity is
via the correlation hierarchy of induced metric fluctuations. This
involves three essential tasks: 1) Deduce the correlations of
metric fluctuations from correlation noise in the matter field;
2) Reconstitution of quantum coherence -- this is the reverse of
decoherence -- from these correlation functions and 3) Use the
Boltzmann-Langevin equations to identify distinct collective
variables depicting recognizable metastable structures in the
kinetic and hydrodynamic regimes of quantum matter fields and how
they demand of the corresponding spacetime counterparts. This
will give us a hierarchy of generalized stochastic equations --
call them the Boltzmann-Einstein hierarchy for quantum gravity --
for each level of spacetime structure, from the macroscopic
(general relativity) through the mesoscopic (stochastic gravity)
to the microscopic (quantum gravity). These equations in the
Boltzmann-Einstein hierarchy should be derivable from a master
effective action of quantum gravity, a good challenge for the
future. \footnote{This is not the Einstein-Boltzmann equation in
classical general relativity and kinetic theory which frames the
classical matter in the Boltzmann style as source of the Einstein
equation. Our quantum or stochastic Boltzmann-Einstein equation
refers to gravity and spacetime alone. `Boltzmann' is to show the
kinetic theory nature, and `Einstein' to show its spacetime
structure, albeit these two giants provide their respective
theories only for the lowest correlation order and its dynamics:
in distribution function of spacetime and in
geometro-hydrodynamics respectively.}

Before we close let me add a remark that another route in
connecting general relativity (read: hydrodynamics) to kinetic
theory is via the covariant Wigner function in curved spacetimes.
This was initiated by Calzetta, Habib, Hu \cite{CHHK,CHWigQC} and
pursued further by Antonsen and Fonarev \cite{AntFon}. It has the
advantage that much work has been done in relating classical to
quantum and relating hydrodynamics to kinetic theory via the
Wigner function. The Wigner function obeys a Fokker-Planck
equation; indeed a Wheeler-De Witt-Vlasov equation for
mini-superspace quantum cosmology has been obtained before
\cite{CHWigQC}. The kinetic theory formulation of quantum gravity
via the Wigner function can be carried out in parallel to the
Einstein-Langevin formulation presented here (if we talk about
pure spacetime structure without quantum fields, we can assume
that the quantum field source is played by gravitons), as the
relation between a Langevin equation and a Fokker-Planck equation
is well-known. There is a hierarchy of Wigner functions
corresponding to that of the correlation hierarchy discussed
here.  What needs to be done in the latter approach to bring it
to a par with the former is to extract the noise term (see, e.g.,
\cite{stobol,NoiseQFT})at the level of stochastic Boltzmann
equation if we work with an interacting quantum field, or with
the Boltzmann-Einstein hierarchy of quantum gravity if we refer
to the hierarchy of the micro-structure of spacetime. The nature
of our quest and the task to be performed are similar.

Now to the tasks: Indications of task 1) is exemplified by the
recent work of Martin and Verdaguer \cite{MV2} where they derived
for perturbations off Minkowski space the Einstein tensor
correlation function. An example of task 3) is the metric
conductance fluctuations explored in the work of Shiokawa
\cite{Shiokawa}. Task 2), the reconstitution of quantum phase
information from correlation functions in the field, is
illustrated here with a suggestion based on the correlation
hierarchy applied to the black hole information `loss' puzzle
\cite{BHinfo}. We intend to return to this issue with the model
of a relativistic particle moving in a quantum field
\cite{HJRinfo} based on the theses works of Philip Johnson
\cite{Johnson} and Alpan Raval \cite{Raval}). Further discussions
of hydrodynamic excitations of spacetimes can be found in
\cite{Sak3,QGBEC}. Last but not least, even though we focus here
on the kinetic / hydrodynamic theory and noise / fluctuations
aspects in seeking a clue to the micro theory of spacetime from
macroscopic constructs, another very important factor is
topology. Topological features can have a better chance to
survive the coarse-graining or effective / emergent processes to
the macro world and can be a powerful key to unravel the
microscopic mysteries.

To conclude, a simple message of this paper is that at our
current level of understanding perhaps it is more fruitful or
even correct to probe into the macro to micro aspect than simply
quantizing the metric or connection form of general relativity.\\

\noindent {\bf Acknowledgement} The ideas expressed here grew
from my thoughts in work done for the last fifteen years, many in
collaboration with Esteban Calzetta. The recent work of Enric
Verdaguer and his students on stochastic gravity provides the
most thorough and solid explication of this theory. I thank them
for many enjoyable discussions and their enduring friendship. I
started learning nonequilibruim statistical mechanics two decades
ago by asking Professor Robert Dorfman questions and sitting in a
course offered by Professor Robert Zwanzig. I'd like to take this
opportunity to express my gratitude to these two esteemed
colleagues. I would also like to thank two other close colleagues
and friends, Ted Jacobson, for sharing some of my views yet
challenging me with probing questions, and Professor Ching-Hung
Woo, with his thoughtful comments. The Peyresq meetings have
provided a forum for these ideas to be explored and aired. I wish
to thank the principal organizer, Edgard Gunzig, for his skillful
organization and warm hospitality. This research is supported in
part by the National Science Foundation under grant PHY98-00967.



\end{document}